# Mesures à large bande et analyse du blocage humain dans un canal radio indoor à 60 GHz

Mbissane Dieng*, Gheorghe Zaharia, Ghaïs El Zein
Univ Rennes, INSA Rennes CNRS, IETR, UMR 6164, F 35000 Rennes, France
Email: {mbissane.dieng ; gheorghe.zaharia ; ghais.el-zein}@insa-rennes.fr

*Résumé* —Les nouveaux systèmes de communication sans fil en bande millimétrique peuvent être fortement impactés par le blocage introduit par le corps humain. A 60 GHz, la couverture de ces systèmes est relativement limitée en raison des pertes de propagation élevées. Ainsi, la formation de faisceaux (beamforming) permet de trouver un chemin réfléchissant pour remplacer celui bloqué. Dans ce travail, l'étude se concentre sur l'impact d'un bloqueur humain dans une salle de réunion, pour évaluer les pertes de blocage introduites par le corps humain à 60 GHz. Les résultats obtenus en termes d'affaiblissement de trajet et de réponse impulsionnelle du canal montrent que l'atténuation par le corps humain est comprise entre 24 et 26 dB. De plus, les résultats montrent que l'utilisation du beamforming permet d'exploiter les trajets réfléchis pour remplacer la liaison directe qui peut être bloquée par le corps humain.

*Mots-clés : ondes millimétriques, beamforming, 60 GHz, blocage humain, propagation indoor*

## I. INTRODUCTION

Dans les futurs systèmes de communication sans fil, les ondes millimétriques (mmWaves) joueront un rôle important pour répondre à des débits de données élevés. Cependant, du fait de leurs courtes longueurs d'onde, ces ondes millimétriques présentent des pertes de propagation élevées et sont très atténuées par le blocage, qui peut être dû au corps humain.

L'une des premières études de mesure de blocage humain [1] a été menée à 60 GHz pour les réseaux locaux sans fil et dans un environnement de type bureau. Des pertes d'environs 20 dB ont été relevées lorsqu'une personne bloquait le chemin direct avec des antennes omnidirectionnelles et de 30 dB en utilisant des antennes directives [2, 3]. Des études récentes ont utilisé différentes approches pour modéliser le blocage dans les bandes mmWave, en tenant compte de différents paramètres de la liaison radio : position Tx-Rx, distance, fréquence, gain d'antenne et hauteur d'antenne.

Dans cet article, nous présentons des mesures de propagation radio à 60 GHz dans une salle de réunion. Les résultats de mesure, portant sur l'affaiblissement de trajet et sur la réponse impulsionnelle du canal, sont analysés d'une part, pour étudier l'impact du blocage par le corps humain et d'autre part, pour étudier la possibilité d'implémenter le beamforming comme solution contre le blocage. Dans la suite, nous allons présenter le système de mesure, l'environnement, le scénario étudié et pour terminer les résultats obtenus.

## II. MESURES DU CANAL INDOOR

### A. Système de mesure

Le système de mesure utilisé est construit autour d'un analyseur de réseau vectoriel (VNA) qui effectue un balayage de fréquence sur une bande de 2 GHz, centrée sur la fréquence intermédiaire (FI) de 3,5 GHz (Fig. 1).

Plus de détails sur le système de mesure sont donnés dans [4].

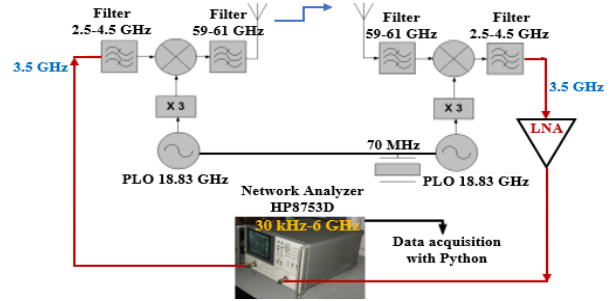

Fig. 1. Système de mesure.

Le tableau I résume les paramètres du système de mesure.

TABLEAU I. PARAMETRES DU SYSTEME DE MESURE

| Fréquence centrale (GHz) | 60 |
|---|---|
| Fréquence intermédiaire (GHz) | 3,5 |
| Bande de mesure (GHz) | 2 |
| Nombre de points de fréquence | 401 |
| Pas de fréquence (MHz) | 5 |
| Puissance de transmission (dBm) | 0 |

Deux antennes cornet ont été utilisées côté Tx et Rx, en polarisation verticale. Elles ont un gain de 22,5 dBi et une ouverture de faisceau à -3 dB de 13° en azimut et 10° en élévation.

### B. Environnement de mesure

Les mesures ont été réalisées dans un environnement statique constitué d'une salle de réunion dont les dimensions sont : 6,5 x 2,2 x 2,5 m$^3$ (Fig. 2).

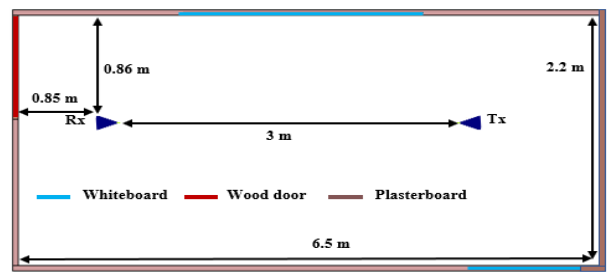

Fig. 2. Environnement de mesure en salle de réunion.

Dans cette pièce, se trouvent deux tableaux blancs de mêmes dimensions, d'une longueur de 2 m et d'une largeur de 1,2 m. Cette pièce possède également une porte en bois de 2,3 m de haut et 0,96 m de large. La porte est restée fermée pendant toute la campagne de mesures. La présence du mobilier dans la salle (chaises, table) a pu être négligée lors de l'analyse des résultats de mesures, du fait de l'utilisation d'antennes très directives et assez élevées.

### C. Scénario de mesure

Les antennes sont séparées d'une distance de 3 m avec 1,71 m de hauteur (Fig. 3).

Un bloqueur humain, dont la largeur aux épaules, l'épaisseur et la hauteur sont respectivement de 45 cm, 13 cm et 1,72 m, a été considéré.

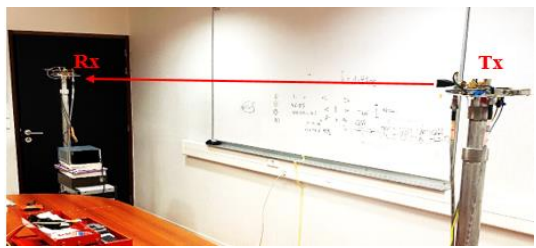

Fig. 3. Scénario de mesure dans la salle de réunion.

Pour cette campagne de mesures, trois scénarios différents sont envisagés :

- Cas où les antennes sont en visibilité directe (LOS, pour Line Of Sight))
- Cas où le chemin direct est bloqué
- Cas avec blocage du chemin direct et dépointage des antennes Tx et Rx de 30° vers le tableau.

### III. Résultats De Mesure

La figure 4 représente l'évolution des pertes de trajet (PL), avec et sans blocage, en fonction des angles d'arrivée.

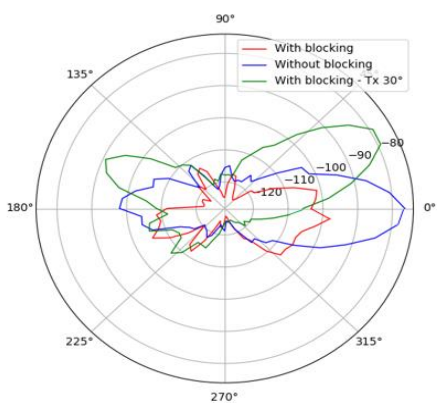

Fig. 4. Affaiblissement de trajet en fonction des angles d'arrivée.

Dans le cas où les antennes sont en visibilité, on obtient un affaiblissement de trajet de 77,2 dB sans blocage et de 103 dB avec blocage humain. Ces résultats permettent d'estimer les pertes introduites par le bloqueur à environ 25,8 dB, ce qui n'est pas loin des résultats présentés dans [2] à 60 GHz. Par ailleurs, le trajet réfléchi provoque une perte de seulement 1,8 dB par rapport au trajet direct. Les figures 5 et 6 montrent les profils puissance relative–retard (PDP), c'est-à-dire le PDP normalisé au niveau de signal reçu maximal sur tous les PDP.

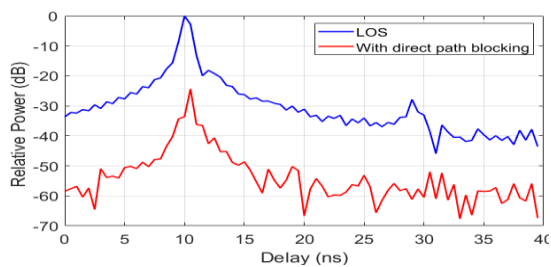

Fig. 5. PDP relatif avec et sans blocage humain.

Dans le cas LOS (Fig. 5, courbe bleue), le trajet direct arrive avec un retard de 10 ns, ce qui correspond bien à la distance de 3 m séparant les antennes Tx et Rx.

En présence d'un bloqueur sur le trajet direct, on obtient une valeur de puissance relative d'environ - 24,4 dB avec un retard de 10,5 ns (courbe rouge). Ce retard correspond à une distance de 3,15 m, soit une différence de marche par rapport au trajet direct de 0,15 m. Ainsi, les pertes de blocage relevées sur les PDP (24,4 dB) sont proches de celles calculées précédemment avec les pertes de trajet, qui sont égales à 25,8 dB.

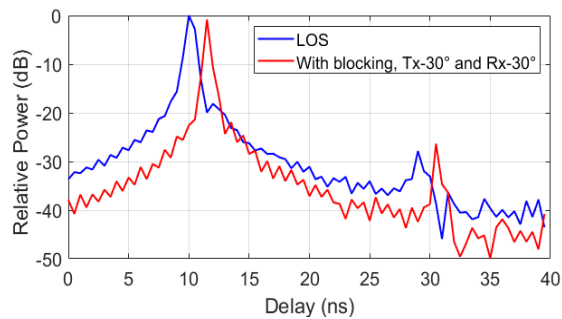

Fig. 6. PDP relatif avec blocage et dépointage des deux antennes de 30°.

Dans le cas où les deux antennes sont pointées vers le centre du tableau (Fig. 6), on obtient un trajet réfléchi avec un retard de 11,5 ns et une puissance relative d'environ -0,84 dB. En comparant le niveau reçu par rapport à celui relatif au trajet direct sans blocage, on constate une légère diminution de 0,84 dB.

### IV. Conclusion

Les résultats obtenus montrent que, lorsque les antennes sont pointées l'une vers l'autre, la présence d'un bloqueur humain introduit des pertes supplémentaires, généralement comprises entre 24 et 26 dB. D'autre part, le calcul de la réponse impulsionnelle du canal permet de mettre en évidence le phénomène de trajets multiples et conduit à l'obtention des PDP pour les différentes configurations de mesure. On note aussi que le trajet réfléchi peut être utilisé lorsque le trajet direct est bloqué, en utilisant à la fois la formation de faisceaux côté émission et réception.

Toutefois, l'environnement considéré ici présente des conditions très favorables aux chemins réfléchis. D'autres campagnes de mesures sont prévues dans des environnements aux conditions de réflexion moins favorables, avec d'autres types d'antennes et de matériaux réfléchissants (murs, portes, fenêtres, etc.).